\documentstyle[11pt]{article}
\newcommand{\al}{\alpha}
\newcommand{\be}{\beta}
\newcommand{\vr}{\vec{r}}
\newcommand{{ \vn }}{\vec{n}}

\newcommand{\beq}{\begin{equation}}
\newcommand{\eneq}{\end{equation}}

\begin{document}

\begin{titlepage}

\large
\baselineskip=16pt


\small{\bf INFN-NA-IV 17/98}

{\Large \bf
\begin{center}
{Study of the critical properties of the Quantum Hall Fluid in
the framework  of a dual statistical model.}
            \footnote{Work supported in part by MURST and by EC contract n.
FMRX-CT96-0045.}
\end{center} }

\begin{center}
G. Cristofano$^{* \dagger}$
 D. Giuliano$^{ \dagger \%}$,
G. Maiella$^{* \dagger}$, F. Nicodemi$^{* \dagger}$

 \vspace{0.2cm}

{\normalsize $^{*}$ Dipartimento di Scienze Fisiche,
Universit\`a di Napoli
\\
$^{\dagger}$ INFN - Sezione di Napoli \\
$\% $ INFM - Sezione di Napoli

Mostra d'Oltremare, Pad.19,  80125 Napoli, Italy}
\end{center}

\vspace{1cm}

{\centerline{\bf Abstract}

By using Renormalization Group methods we analyze the outcome
of a recent proposal of
describing the Quantum Hall fluid in terms of a dual
plasma which embodies dyons as effective degrees of freedom.
As a consequence the
 physical interpretation of the two parameters of the model
as the longitudinal and the Hall conductances is made clear.
Parameters' scaling roperties
allow for the determination of the critical index
for the localization length after a mapping
of our statistical model onto a classical percolation model. The
universality of the critical properties, supported by recent
experiments, is a consequence of the infinite discrete symmetry of the
model $SL(2, Z)$ (generalized duality), as noticed in a previous
paper.}

\vspace{1cm}

\hfill\begin{center} PACS numbers:  73.40.Hm; 05.70.Fh; 11.10.Kk \end{center}

\vspace{1cm}


\end{titlepage}

\newpage

\section{Introduction.}

Several years ago it was discovered that the ground-state wave
function  proposed by Laughlin \cite{lau} for a Hall system at filling
$f = 1/m$, where $m$ is an odd integer,
is simply described (at least for its analytic part) in terms of
correlators of primary fields of a two-dimensional Conformal Field
Theory ({\bf 2D CFT}) with central charge $c = 1 $ \cite{fub, cmmn}.

Within such a framework a more straightforward and physical picture of
the Laughlin state was given in terms of effective degrees of freedom,
which play the role of ``collective modes" of the otherwise strongly
correlated electrons system.
Laughlin's physical idea of associating a magnetic flux to the electron
\cite{lau} is in agreement with a detailed analysis of the ground state
properties of the Hall system with doubly periodic boundary conditions
\cite{cmmn2}. More specifically for filling $f = 1/m$ the allowed
magnetic translations have a finite step which defines an ``elementary
cell" for the electron with a linked magnetic flux just equal to $m
\Phi_0$, where the ``elementary flux" $ \Phi_0$ is expressed in terms of
universal quantities as $\Phi_0 = \frac{hc}{e}$. The previous
observation suggests a picture of the Hall fluid at a given filling in
terms of ``dyons" (i.e., objects carrying both electric and magnetic
charge) as the relevant degrees of freedom.
Then the conduction properties are simply reproduced by employing the
transformation properties of the ground state under the finite magnetic
translations and its topological properties are easily derived.

Furthermore the phenomenological ``laws of corresponding states"
\cite{klz} suggest an unified description of the Hall fluid at
integer as well as at fractional filling in terms of a discrete symmetry.
Such a symmetry has been used by several authors \cite{klz, jain} but
only on a qualitative ground.

In this paper we analyze the consequences of a recent proposal of
describing the Quantum Hall fluid which incorporates in a simple way both
the phenomenon of dyon condensation and the discrete symmetry $SL(2, Z)$
\cite{nostro}.
By employing simple Renormalization Group ( {\bf RG} ) techniques we are able to study
the flow of the parameters appearing in the model. As a result we are let
to identify the RG Infrared ( {\bf IR} ) attractive fixed points as stable condensates
of the Hall fluid at the plateaux. The $SL(2, Z)$ symmetry acts as in
\cite{ca, cara}, mapping the IR fixed points into one  another (a
simple connection between the duality transformations of the model and the
laws of corresponding states has been given in
\cite{nostro}). Furthermore we study the properties of the model around the RG
repulsive fixed points, once identified as transition points between
plateaux. After mapping our model onto a classical percolation model, the
scaling properties of the longitudinal and Hall conductances (corresponding
to the two parameters appearing in the model) allow us to fix the critical
index for the localization length at the value $\nu = \frac{4}{3}$.
Universality of the result again is assured by the discrete $SL(2,
Z)$ symmetry, mapping repulsive fixed points into one another.

For clarity sake we should make an obvious comment:
our two-dimensional model is suitable for a description of
the equilibrium properties of the Hall system.

The paper is organized as follows:

\begin{itemize}

\item In section 2 we review the construction of
a non-trivial formulation of the
Cardy-Rabinovici ( {\bf CR}) model where
we split the field configuration in a uniform ``background" part
plus a fluctuating field in order to describe a Coulomb Gas of charges
of the same sign neutralized by a classical background of opposite
sign \cite{nostro}.
We observe the existence of an infinite discrete
symmetry $SL(2,Z)$.

\item Section 3 contains a detailed RG analysis
of the model which
allows for a simple description of its long distance properties
 and for a physical
interpretation of the parameters appearing in the model.
As outcome  we find that there are stable phases
corresponding to IR RG fixed points.
The $SL( 2, Z)$ symmetry maps those
non-trivial fixed points into one another
suggesting a unified
picture of them in terms of a 2D CFT
 with central charge $c=1$.

\item In section 4 our model, taken at the transition point, is mapped onto
a percolation model and the critical index relative to the scaling of the
localization length is derived.

\item In section 5 we give a summary of the results and
address some open questions.

\item In the Appendix some properties of the $SL(2, Z)$ fixed points are
discussed.

\end{itemize}

\section{Construction and properties of the dual plasma model.}

\subsection{Cardy-Rabinovici model.}

Let us  first remind how Cardy and Rabinovici build their 2D model
starting from a $U(1)$ gauge theory with both electric and magnetic
matter coupled by a $\theta$ term \cite{ca,cara}. Then we will search for
a non trivial extension of such a model in which a fixed background
is generated.

The explicit Euclidean action is
given by:
\beq
S [ A^j, \vec{S}^j,  n^j ] \equiv \frac{1}{2g}\int d^2r\sum_{ j = 3, 4}
(\partial_\beta A^j +
S^j_\beta)^2 - i\int d^2 r \sum_{ j = 3, 4} n^j A^j
 - i\frac{\theta}{2\pi}\int d^2r
\epsilon^{ij} \epsilon_{\beta\gamma}S^i_\beta \partial_\gamma A^j ~~,
\label{start}
\eneq

\[
( j = 3, 4 )
\]

where $\epsilon$ is the
antisymmetric tensor in 2D,
$A^j$ are a pair of scalar fields and
$S^j_\beta(\vr)$ are the ``magnetic frustration fields" satisfying
the constraints:

\[
\epsilon_{\beta\gamma}\partial_\beta S^3_\gamma (\vr) - m^4(\vr) = 0
\]

\beq
\epsilon_{\beta\gamma}\partial_\beta S^4_\gamma (\vr) + m^3(\vr) = 0
\label{const}
\eneq

while $n^j (\vr)$ and $m^j (\vr)$ are  the electric and
magnetic charge densities.

The densities ($n^j$, $m^j$)
are constrained by the neutrality condition (required
to make the system IR stable):

\beq
\int d^2 r~ n^j (\vr) = \int d^2 r ~m^j (\vr) = 0 ~~.
\label{neut}
\eneq

The Coulomb gas representation,  where the role of the electric and
magnetic charges is emphasized, is defined by:

\beq
e^{-S_{CG}(n^j, m^j)} = \int \prod_{j = 3, 4}
 {\cal D}A^j \int \prod_{\alpha=1,2} {\cal
D}S^j_\alpha \delta (\epsilon_{\beta\gamma}\partial_\beta S^j_\gamma -
\epsilon^{ij} m^i)
e^{- S [ A^j, \vec{S}^j , n^j]}~~.
\label{cogar}
\eneq

It is straightforward to evaluate the path integrals in eq.( \ref{cogar} ) and
obtain:

\begin{eqnarray}
S_{CG} [ n^j, m^j] = \frac{g}{2}\int d^2r d^2r^{'} \sum_{j =3, 4}
(n^j(\vec{r}) +
\frac{\theta}{2\pi} m^j(\vec{r})) (n^j(\vec{r}^{'}) +
\frac{\theta}{2\pi} m^j(\vec{r}^{'})) G(\vec{r}-\vec{r}^{'})+\\
\nonumber
\frac{1}{2g}\int d^2r d^2r^{'}\sum_{j =3, 4}
 m^j(\vec{r})  m^j(\vec{r}^{'}) G(\vec{r}-\vec{r}^{'})+
i \int d^2r d^2r^{'}\epsilon^{ij} n^i(\vec{r})
m^j(\vec{r}^{'})\varphi (\vec{r} -
\vec{r}^{'})~~,
\label{cgrep}
\end{eqnarray}

where $G(\vec{r})$ and $\varphi(\vec{r})$ are the ``longitudinal" and
``transverse" Green-Feynman functions in 2D given by:
\beq
G(\vec{r}) = \ln\left(\frac{|\vr|}{a}\right) ~~,~ \varphi(\vec{r}) = \arctan
\left(\frac{y}{x}\right) ~~,
\eneq
here $a$ is a cutoff.
The last term in eq.( \ref{cgrep} ) is the (imaginary) Bohm-Aharonov
term. Also notice that for $ \theta = 0$ the standard
Coulomb gas for both electric and magnetic charges is reproduced
( see \cite{nie}).

\subsection{The model with harmonic background.}

To obtain a magnetic
charge distribution consistent with Laughlin's description of the
Hall fluid, i.e. the one-component plasma where all the vortices have the
same magnetic charges and interact with an uniform external background
of opposite sign which
exactly neutralizes  their total magnetic charge \cite{lau}, we have
to modify the CR model.

As a first step we introduce the background in the model
described by the action of eq.( \ref{start})
which, for $\theta = 0$, is given by:

\beq
S = \frac{1}{2g}\sum_{j=3, 4} \int d^2 r ( \vec{\nabla} A^j
 + \vec{S}^j )^2 - i \int d^2 r n^j A^j
\eneq

where $\vec{S}^j$ must obey the constraint given by eq.( \ref{const} ).

The next step is the splitting of the charge densities into an
uniform and a variable term:

\beq
m^j ( \vr )  = \bar{m}^j + \mu^j ( \vr ) \;\;\;
n^j ( \vr )  = \bar{n}^j + \nu^j ( \vr )
\eneq

and the neutrality condition of eq.( \ref{neut} ) can be now rewritten as:

\beq
\int d^2 r \mu^j ( \vr ) + \bar{M}_j = 0 \;\;\; ; \;\;
\int d^2 r \nu^j ( \vr ) + \bar{N}_j = 0
\label{nneut}
\eneq

where $\bar{M}^j = A \bar{m}^j$ and $\bar{N}^j = A \bar{n}^j$, $A$ being the
area of the sample and integration over the whole sample is understood.

Also by defining:

\[
\vec{S}^j ( \vr ) = \vec{\bar{S}}^j ( \vr ) + \vec{\sigma}^j ( \vr )
\]

where the uniform part $\bar{S}^j$  obeys the equation

\[
\epsilon_{\beta \gamma} \partial_\beta \bar{S}_\gamma^j ( \vr ) =
\epsilon^{  j k} \bar{m}^k
\]

and $\sigma ( \vr )$ is the fluctuating part, we obtain:

\[
\frac{1}{2g} \sum_{j = 3, 4} \int d^2 r ( \vec{\nabla} A^j + \vec{S}^j )^2 =
\]

\beq
\frac{1}{2g}\sum_{j = 3, 4}
 \int d^2 r ( \vec{\nabla} A^j + \vec{\sigma}^j )^2
+ \frac{1}{g} \int d^2 r \vec{\bar{S}^j} \cdot
( \vec{\nabla} A^j + \vec{\sigma}^j )
+ \frac{1}{2g} \int d^2 r ( \vec{\bar{S}}^j )^2
\label{elma}
\eneq

The integral $\frac{1}{2g} \int d^2 r
( \vec{\bar{S}}^j )^2$  is
a simple number independent of the field configuration we
shall disregard it. Furthermore the term
$ \frac{1}{g} \int d^2 r \vec{\bar{S}}^j \cdot  ( \vec{\nabla} A^j )$
will be zero once we fixed $\vec{\bar{S}^j}$ in such a way that $\vec{\nabla}
\cdot \vec{\bar{S}}^j = 0$.

The term $ \frac{1}{g} \int d^2 r \vec{\bar{S}}^j \cdot \vec{\sigma}^j$
gives instead:

\[
- \sum_{j = 3, 4} \frac{\bar{m}^j}{4 g} \int d^2 r \mu^j ( \vr ) r^2
\]

In order to work out the term:

\[
\tilde{S} = \frac{1}{2 g} \sum_{j = 3, 4} \int d^2 r ( \vec{\nabla} A^j +
\vec{\sigma}^j)^2
- i\sum_{j = 3, 4} \int d^2 r ( \bar{n}^j + \nu^j ) A^j
\]

we define $A^j = a^j + \bar{A}^j$, where $\bar{A}^j$ is the
background field satisfying:

\beq
\nabla^2 \bar{A}^j + i g \bar{n}^j = 0
\eneq

After imposing isotropic boundary conditions for $\bar{A}^j$ we get:

\beq
\bar{A}^j = - i g\frac{\bar{n}^j}{4} r^2
\eneq

Then the action in the Coulomb Gas representation can be written as
follows:

\[
S_{CG} = - \sum_{j = 3, 4}
\frac{1}{2} \int d^2 r  \int d^2 r^{'} \left[ \frac{\mu^j (
\vr ) \mu^j ( \vr^{'} )}{g} + g \nu^j ( \vr ) \nu^j ( \vr^{'} ) \right]
 G (\vr - \vr^{'} )
-
\]

\[
i \epsilon^{ij}\int d^2 r  \int d^2 r^{'} \mu^i ( \vr ) \nu^j
 ( \vr^{'} ) \varphi ( \vr - \vr^{'} ) -
\sum_{j = 3, 4} \frac{\bar{m}^j}{4 g} \int d^2 r  \mu^j ( \vr ) r^2 -
\]

\beq
\frac{g \bar{n}^j}{4} \int d^2 r  \nu^j ( \vr ) r^2
\eneq

Defining for the variable part of the charge densities:

\beq
\mu^j ( \vr ) = \sum_{i = 1}^N \mu_i^j \delta ( \vr - \vr_i )
\;\;\;
\nu^j ( \vr ) = \sum_{i = 1}^N \nu_i^j \delta ( \vr - \vr_i )
\label{chd}
\eneq

we finally get for the partition function of $N$ particles in the
presence of a background:

\[
Z_N  = \int \prod_{k = 1}^N \frac{d^2 r_k }{a^2} \exp \{
\frac{1}{2} \sum_{j = 3, 4}
\sum_{i \neq k = 1}^N \left[ \left( \frac{ \mu^j_i \mu^j_k}{g}
+ g \nu^j_i \nu^j_k \right) \ln \left| \frac{\vr_i - \vr_k}{a} \right|
\right]
\]

\[
+ i \epsilon^{rj}
\sum_{i \neq k =1}^N \mu^r_i \nu^j_k \varphi ( \vr_i - \vr_k )
\} \times
\]

\beq
\exp \left\{ \frac{1}{4} \sum_{j = 3, 4}
\left[ \frac{\bar{m}^j}{g} \sum_{k  1}^N \mu^j_k
r_k^2 + g \bar{n}^j \sum_{k = 1}^N \nu_k^j r^2_k \right] \right\}
\eneq

The presence of a $\theta$-term introduces an interaction between electric
and magnetic charges obtained by the usual substitution:

\[
\nu_i^j \rightarrow \nu_i^j + \frac{\theta}{2 \pi} \mu_i^j
\]

Therefore the complete form for the partition function for $N$
particles is given by:

\[
Z_N  = \int \prod_{k = 1}^N \frac{d^2 r_k }{a^2} \exp \{
\frac{1}{2} \sum_{j = 3, 4}
\sum_{i \neq k = 1}^N \left[ \left( \frac{ \mu^j_i \mu^j_k}{g}
+ g ( \nu^j_i + \frac{\theta}{2\pi} \mu^j_i )
 ( \nu^j_k + \frac{\theta}{2\pi} \mu^j_k )
  \right) \ln \left| \frac{\vr_i - \vr_k}{a} \right|
\right]
\]

\[
+ i \epsilon^{rj}
\sum_{i \neq k =1}^N \mu^r_i ( \nu^j_k +\frac{\theta}{2\pi} \mu^j_k )
\varphi ( \vr_i - \vr_k )
\} \times
\]

\beq
\exp \left\{ \frac{1}{4} \sum_{j = 3, 4}
\left[ \frac{\bar{m}^j}{g} \sum_{k =  1}^N \mu^j_k
r_k^2 + g  ( \bar{n}^j + \frac{\theta}{2\pi} \bar{m}^j )
\sum_{k = 1}^N ( \nu^j_k +\frac{\theta}{2\pi} \mu_k^j )
r^2_k \right] \right\}
\label{compl}
\eneq

and the neutrality condition given by eq.( \ref{nneut} ), with the help
of eq.( \ref{chd} ) can be written as:

\beq
\sum_{i = 1}^N \mu_i^j + \bar{M}^j = 0 \;\;\; ; \;\;
\sum_{i = 1}^N \nu_i^j + \bar{N}^j = 0
\label{newne}
\eneq

In eq.( \ref{compl} ) we notice the presence of a
 harmonic  background term which has the correct form in order to describe the
Hall fluid at the plateaux \cite{lau}.
Details can be found in \cite{nostro}.

\subsection{Discrete symmetries of the model: the modular group $SL(2,
Z)$.}

We now study the discrete symmetries of our model.
It has been already noticed that the introduction of the $\theta$ angle
allows for an extension of the usual electric-magnetic duality to a
generalized discrete non-abelian symmetry which we shall refer to as
generalized duality and which is described by the modular group
$SL( 2 , Z)$ \cite{ca, cara} that acts on the complete action of the model,
eq.( \ref{compl} ).

In order to study the symmetries of the model we have to look at the
explicit form of the action. We then easily find the following
symmetry transformations:

\begin{itemize}

\item Periodicity $\hat{T}$:

\[
\frac{\theta}{2\pi} \rightarrow \frac{\theta}{2\pi} + 1 \;\;\; ; \;\;
\frac{1}{g} \rightarrow \frac{1}{g}
\]

\[
\bar{n}^j \rightarrow \bar{n}^j - \bar{m}^j
\;\;\; ; \;\;
\bar{m}^j \rightarrow \bar{m}^j
\]

\beq
\{ \nu^j \} \rightarrow \{ \nu^j \} - \{ \mu^j \}
\;\;\; ;\;\; \{ \mu^j \} \rightarrow \{ \mu^j \}
\label{perio}
\eneq

\item Duality $\hat{S}$:

\[
\frac{1}{g} \rightarrow \frac{\frac{1}{g}}{ \left( \frac{1}{g} \right)^2
+ \left( \frac{\theta}{2\pi} \right)^2 }
\;\;\; ; \;\;
\frac{\theta}{ 2 \pi} \rightarrow \frac{- \frac{\theta}{2\pi}}
{ \left( \frac{1}{g} \right)^2
+ \left( \frac{\theta}{2\pi} \right)^2 }
\]

\[
\bar{n}^j \rightarrow \bar{m}^j \;\;\; ; \;\;
\bar{m}^j \rightarrow - \bar{n}^j
\]

\beq
\{ \nu^j \} \rightarrow \{ \mu^j \} \;\;\; ; \;\; \{ \mu^j \}
\rightarrow - \{ \nu^j \}
\label{mod}
\eneq

\end{itemize}

(Notice that, in order to define symmetries of our model, we have to make
the transformations to act on the charge densities as well as on the
background.)

The above discrete transformations are more easily described
in terms of the complex parameter:

\[
\zeta \equiv \frac{\theta}{2\pi} + i \frac{1}{g}
\]

on which the  transformations $\hat{S}$, $\hat{T}$ act as follows:

\[
\hat{T} \;\; : \;\; \zeta \rightarrow \zeta + 1
\]

\beq
\hat{S} \;\; : \;\; \zeta \rightarrow - \frac{1}{\zeta}
\eneq

Then $\hat{S}$ and
$\hat{T}$ generate the group ${\cal G}$ which acts on $\zeta$ as:

\beq
{\cal G} \;\; : \;\; \zeta \rightarrow \frac{A \zeta + B}{C \zeta + D}
\eneq

where $A,B,C,D, \in Z$ and $AD - BC = 1$, defining the discrete group
$SL(2, Z)$, relevant for the analysis of the Hall plateaux hierarchy
\cite{jain, zhk}.

In the Appendix some properties of the $SL(2, Z)$ fixed points are
discussed.

\section{Renormalization Group analysis.}

\subsection{Linear approximation in the fugacities.}

We are now ready to derive the RG equations in the linear (in the
fugacities) approximation  and in the presence of a non-trivial
background term.

Since the RG analysis is performed by rescaling the cut-off $a$, the
Bohm-Aharonov term will be left unchanged and correspondingly the theory
gets factorized with respect to the internal indices 3, 4. So for the moment
we drop the internal indices and perform the linear RG analysis
using  the simplified version of the action:

\[
S = \frac{1}{2} \sum_{i \neq k = 1}^N \left[ \frac{ \mu_i \mu_k}{g}
+ g \left( \nu_i + \frac{\theta}{2\pi} \mu_i \right) \left( \nu_k
+ \frac{\theta}{2 \pi} \mu_k \right) \right]
\ln \left| \frac{\vr_i - \vr_k}{a} \right|
+
\]

\beq
 \frac{1}{4} \sum_{i = 1}^N \left[
\frac{\bar{M} \mu_i}{g A} + \frac{g}{A}
\left( \nu_i + \frac{\theta}{2 \pi} \mu_i \right)
\left( \bar{N} + \frac{\theta}{2 \pi} \bar{M} \right) \right] r_i^2
\label{simpl}
\eneq

By rescaling $a$ as:
$a \rightarrow a ( 1 + \lambda)$ ($\lambda =
\frac{da}{a}$)
the action $S$ gets transformed as follows:

\[ S \rightarrow S - \frac{ \lambda}{2}
\sum_{i \neq j = 1}^N \left[ \frac{\mu_i
\mu_j}{g} + g \left( \nu_i + \frac{\theta}{2\pi }\mu_i \right) \left( \nu_j
+ \frac{\theta}{2\pi} \mu_j \right) \right] = \]

\beq S +\frac{ \lambda}{2}
 \sum_{j = 1}^{N} \left[ \frac{( \bar{M} + \mu_j) \mu_j}{g}
+ g \left( \bar{N}+ \nu_j +\frac{\theta}{2\pi} (\bar{M}+\mu_j) \right)
\left( \nu_j + \frac{\theta}{2\pi} \mu_j \right) \right] \eneq

where the  neutrality conditions  (see
eq.( \ref{newne} ))
have been taken into account.

Being the integration measure for each particle $\frac{d^2 r_j}{a^2}$,
we find that the change in the Partition Function may be reabsorbed in
terms of the following correction to the fugacities:

\beq Y(\nu , \mu ) \rightarrow Y(\nu, \mu ) + d Y(\nu , \mu) \eneq

where:

\beq d Y( \nu, \mu ) = \lambda \left[2
+ \frac{1}{2} \left(  \frac{(\bar{M} + \mu ) \mu }{g} + g
\left( \bar{N} + \nu + \frac{\theta}{2\pi} ( \bar{M} + \mu ) \right) \left(
\nu + \frac{\theta}{2\pi} \mu \right) \right) \right] Y( \nu , \mu ) \eneq

The scaling index for the fugacity is then given by:

\beq x ( \nu , \mu ) = 2 + \frac{( \bar{M} + \mu ) \mu }{2 g} +
\frac{g}{2} \left(
\bar{N} + \nu  + \frac{\theta}{2\pi} ( \bar{M} + \mu ) \right) \left( \nu +
\frac{\theta}{2\pi} \mu \right)  \eneq

We shall now assume that the
charge which condense will correspond to the maximum possible value of
the scaling indices, for a fixed background and particle number.

That is, by taking into account the constraints expressed by
eq.( \ref{newne} ), we have to maximize

\[ \xi( \nu, \mu ) = x( \nu , \mu ) + \al  ( \sum_{j = 1}^{N} \nu_j +
\bar{N} ) + \be ( \sum_{j = 1}^{N} \mu_j + \bar{M} ) \]

where $\alpha$ and $\beta$ are Lagrange multipliers.

We find:

\[ \al = - \frac{N - 2}{N} g \left[ \bar{N} + \frac{\theta}{2\pi} \bar{M}
\right] \]

\beq \be = - \frac{N - 2}{N} \left[ \bar{M} \left(\frac{1}{g} + g \left(
\frac{\theta}{2\pi} \right)^2 \right) + g \frac{\theta}{2\pi} \bar{N}
\right] \eneq

and the maximum
is reached for the following values of the ``condensing'' dyons:

\[ \nu_1 = \ldots = \nu_{N} = - \frac{\bar{N}}{N} \equiv \nu \]

\beq \mu_1 = \ldots = \mu_{N} = - \frac{\bar{M}}{N} \equiv \mu
\label{crch} \eneq

From the above equations we see that the condensate is made out of
$N$ dyons  whose charges
are $1/N$ times the corresponding backgrounds.
In particular for filling $ f = \frac{1}{m} $, we have:

\[
f \equiv \frac{ \# \; {\rm of \; electrons}}{\# \; {\rm of \; available \;
states}}
= \frac{ \bar{N}}{ \bar{M}} \frac{ h c }{e^2} = \frac{\bar{N}}{\bar{M}}
\]

( being $ h = c = e =1$ in our units)
and by using eq.( \ref{crch} ) we get:

\[
\nu = 1 \;\;\;  ; \;\; \mu = m
\]

Although the result in eq.( \ref{crch} )
looks independent of the initial values of the parameters $g$ and
$\theta$, the latter are important for the stability of the
plateaux, as it will be seen in the next section in the context of
the non-linear approximation of the RG
equations.

We then see that the most relevant consequence of the introduction of a
background is the presence of condensate phases where the charges are all
equal and with the same sign, embedded in an uniform neutralizing
background. Such phases are simply described in terms of the Coulomb
Gas action:

\[ S_N = \left[ \frac{\mu^2}{g} + g \left( \nu + \frac{\theta}{2\pi} \mu
\right)^2 \right] \sum_{i \neq k = 1 }^N \ln \left| \frac{\vec{r}_i -
\vec{r}_k}{a} \right|  \]

\begin{equation}
- \frac{1}{2} \left[ \frac{\mu^2}{g} + g \left( \nu + \frac{\theta}{2\pi}
\mu \right)^2 \right]
\left[ \frac{\bar{m}^2}{g} + g \left( \bar{n} + \frac{\theta}{2\pi}
\bar{m} \right)^2 \right]
 \sum_{k = 1}^N | \vec{r}_k |^2
\label{laga}
\end{equation}

which is a generalization of Laughlin's plasma description.
For a more detailed discussion about this point see
\cite{nostro}.

\subsection{Non linear terms.}

We can now work on the consistency of our
model by studying the flow of the parameters $g$ and $\theta$, induced
by non-linear corrections to the RG equations.

Our technique is a simplified  version of the one
used in \cite{nie}. The basic idea is to begin with a configuration
of the system, described by the action $S_N$,
in which $N$ equal charges have condensed and to perturb the
system around such a configuration by creating a pair of charges equal
in modulus but with opposite signs.

Let $ ( \nu, \mu)$and  $ (-  \nu, - \mu)$ be the charges created with
fugacities
$Y  ( \nu, \mu)$  and $ Y ( - \nu, - \mu)$ and let $Y^2_P$ be the
fugacity corresponding to the particle-antiparticle pair:

\beq
Y^2_P = Y  ( \nu, \mu) Y  ( - \nu, - \mu)
\eneq

When we vary the scale $a$,  $Y^2_P$ scales with an exponent given by:

\[
x_{Y^2_P} =  x   ( \nu, \mu) + x   ( - \nu, -\mu)
= 2 \left[ 2 - \frac{\mu^2}{2g} - \frac{g}{2} \left( \nu +
\frac{\theta}{2\pi}
\mu \right)^2 \right]
\]

From the above equation it is evident that the
values of the ``bare" parameters (i.e., of
the parameters at the scale at which we started)
for which the process of creation
of a pair is relevant must satisfy the condition:

\beq
2 - \frac{\mu^2}{2g} - \frac{g}{2} \left( \nu + \frac{\theta}{2\pi}
\mu \right)^2  > 0
\eneq

This defines a region in the plane of the parameters $( \frac{1}{g},
\frac{\theta}{2 \pi} )$ (see figure), which exactly
coincides with  the one found in \cite{ca,cara} for the case in
which there is no background at all
(we refer to \cite{ca,cara} for a
detailed discussion about the shape of the phase diagram), but
the original model is characterized by phases where the
condensates of charges have  both signs, i.e., they
are trivial from a RG point of view.
For a
Hall system  the condensate is  made of charges with the same
sign whose total charge is neutralized by an uniform background, i.e.,
the model is chiral. Our model allows for a description
 consistent with the physics of the Hall condensates
\cite{lau}, as it will be shortly shown. The relevant observation is that
the phase diagram derived in \cite{ca,cara} continues to be valid in
our model, but
only if applied to (neutral) excitations around the Hall condensate.

In the action
for the $N+ 2$ particles,
$S_{N+2}$, the scale $a$ is sent into $a + d a$. At this
point we have to sum over the configurations in which the centers of
the charges in the pair are at a distance between $a$ and $a + d a$
(``particle fusion") The final result gives an action which is the
same as the one for $N$ particles but with the parameters $g$ and $\theta$
renormalized. In this way we derive the RG equations for the
parameters.

The action for $N + 2$ particles is:

\[
S_{N+2} = \left[ \frac{\mu^2}{g} + g \left( \nu + \frac{\theta}{2\pi} \mu
\right)^2 \right] \{ \sum_{i \neq j = 1 }^N \ln \left|
\frac{\vr_i - \vr_j}{a} \right| + \sum_{j = 1}^N \ln \left| \frac{\vr_j -
\vr_+}{a} \right|
\]

\[
 - \sum_{j = 1}^N \ln \left| \frac{\vr_j - \vr_-}{a}
\right| \} -
\left[ \frac{\mu^2}{g} + g \left( \nu + \frac{\theta}{2\pi} \mu
\right)^2 \right] \ln \left| \frac{\vr_+ - \vr_-}{a} \right|
\]

\beq
- \frac{1}{2} \left[ \frac{\mu^2}{g} + g \left( \nu + \frac{\theta}{2\pi}
\mu \right)^2 \right]
\left[ \frac{\bar{m}^2}{g} + g \left( \bar{n} + \frac{\theta}{2\pi}
\bar{m} \right)^2 \right]
  \left[ \sum_{j = 1}^N \frac{ | \vr_j |^2}{A}
+ \frac{ | \vr_+|^2}{A} - \frac{ | \vr_-|^2}{A} \right]
\eneq

where  $\vr_+$ and $\vr_-$
denote the location of the positive and negative charge respectively.

We rescale  $a$ into $a + da$, with $da = \lambda a$ and $\lambda \ll 1$.
Let us define
$\vec{s} = \vr_+ - \vr_-$ and $\vec{R} = ( \vr_+ + \vr_-)/2$ and sum
over the configurations with $a < | \vec{s}| < a + da$.

The partial sum in the partition function will be given by:

\beq
Y^2_P \int_{|\vec{s}| = a}^{ | \vec{s}| = a + da} \frac{d^2s}{a^2}
\int \frac{d^2R}{a^2} e^{S_{N+2}}
\eneq

where $S_{N + 2}$ is understood to be expressed in terms of the new
variables $\vec{s}$ and $\vec{R}$.

By expanding to second order in $\vec{s}/a$ we get:

\[
\int \frac{d^2 s}{a^2} \left| \frac{ \vec{s}}{a} \right|^{\alpha_q^2}
e^{S_N} \left\{ 1 + \frac{ ( \alpha_q^2)^2}{2} s^a s^b
\frac{\partial \psi ( \vec{R} ) }{\partial R^a}
\frac{\partial \psi ( \vec{R} ) }{\partial R^b}  \right\}
\]

where:

\[
\alpha_q^2 = \frac{\mu^2}{g} + g \left( \nu +\frac{\theta}{2 \pi } \mu
\right)^2
\]

and:

\[
\psi( \vec{R} ) = \sum_{j = 1}^N \ln \left| \frac{\vr_j - \vec{R}}{a} \right|
- \frac{\vec{R}^2}{2A}
\]

We finally get, after integrating:

\[
e^{S_N} + \int \frac{d^2s}{a^2} \frac{d^2 R }{a^2} Y_P^2 e^{S_{N+2}}
\approx
\]

\beq
e^{S_N} \left\{ 1 - Y_P^2 \frac{\pi \lambda}{4} ( \alpha_q^2)^2
\left[ \sum_{i \neq j = 1}^N \ln \left| \frac{\vr_i - \vr_j}{a} \right|
-\sum_{j = 1}^N \frac{ | \vr_j |^2}{2A} \right] \right\}
\eneq

As a consequence $\alpha_q$ gets renormalized according to:

\[
\alpha_q^2 \rightarrow \alpha_q^2 + \delta \alpha_q^2
\]

where

\begin{equation}
\delta \alpha_q^2 = - Y_P^2 \frac{\pi \lambda}{4} ( \alpha_q^2)^2
\end{equation}

Defining the scale parameter $\rho$ through the relation:

\begin{equation}
d \rho = \frac{ \pi \lambda}{4} Y_P^2 = \frac{ \pi }{4} Y_P^2 \frac{da}{a}
\end{equation}

we find the following differential equation:

\begin{equation}
\frac{d \alpha_q^2}{d \rho} = - ( \alpha_q^2 )^2
\end{equation}

with the solution given by:

\beq
\alpha_q^2 ( \rho ) = \frac{ \alpha_q^2 ( \rho_0 )}{ 1 +
\alpha_q^2 ( \rho_0 ) ( \rho - \rho_0 )}
\label{soldif}
\end{equation}

from which we obtain:

\begin{equation}
\lim_{\rho \rightarrow \infty } \alpha_q^2 ( \rho ) = 0
\end{equation}

Being $\alpha_q^2$ a positive-definite quadratic form, the above equation
implies for the parameters $\frac{1}{g}$ and $\frac{\theta}{ 2 \pi}$.

\[
\lim_{\rho \rightarrow \infty} \frac{1}{g ( \rho ) } = 0
\]

\begin{equation}
\lim_{\rho \rightarrow \infty} \frac{\theta ( \rho ) }{2 \pi} = - \frac{\nu}{\mu}
\label{rgpar}
\end{equation}

Notice that the modular covariance of the RG results so obtained simply
relies on the fact that $\alpha_q^2$ is a modular invariant object.

\subsection{Comments about the solutions of the Renormalization Group
equations in the non-linear approximation}

The results expressed in eq.( \ref{rgpar} ) not only are in agreement with
the picture we presented in section 3.1 according to which the most probable
vacuum condensate in our model describes the Laughlin plasma at the
plateaux for a Quantum Hall fluid, but gives also a consistent physical
interpretation of the parameters $\frac{1}{g}$ and $\frac{\theta}{2 \pi} $
appearing in the model. In fact

\[
\lim_{\rho \rightarrow \infty} \frac{1}{g ( \rho )} = 0
\]

corresponds to
the absence of longitudinal conductance in a Hall system in
the thermodynamic limit and $\frac{1}{g ( \rho )}$ is naturally interpreted
as the bare
longitudinal conductance, $\sigma_L$.

On the other hand from eq.( \ref{rgpar} )
the parameter $\frac{\theta (\rho ) }{2\pi}$
can be interpreted as the bare
transverse (Hall) conductance. Infact if
in the thermodynamic limit the
condensate is made out of \underline{dyons} with electric charge $\nu$ and
magnetic charge $\mu$, the transverse conductance $\sigma_H$
will be equal to the ratio
between the two charges, as it must be for a Hall system \cite{nostro}.
Then
the I.R. fixed points of $SL(2, Z)$ in which certain phases condense
are interpreted as points
corresponding to a Hall system at different fillings, whose transverse
conductance is given by the ratio $\frac{\nu}{\mu}$.

Also let us remind that in a real system the values of the physical parameters
are understood to be evaluated at a scale corresponding to the size of the
system. For a finite system, if the size is smaller than the localization
length $\xi$, the bare Hall conductance (i.e., the conductance
evaluated at the size of the system), will be proportional to the reciprocal
of the external magnetic field $B$. This means that a change in the external
field is equivalent in our model to a horizontal variation (at constant $
g ( \rho ) $) of the bare parameter $\frac{\theta ( \rho ) }{2\pi}$
\cite{lea, pru}.
Such an observation gives us an immediate interpretation of the phase
boundaries as ``boundaries" of the plateaux which at fixed disorder has a
size given by the horizontal distance between the phase boundaries enclosing
it (see figure).

Obviously the experimental results have to be compared with the values of
the parameters in the thermodynamic limit, which does not depend on the
initial point, if the phase does not change. But it is also important to
give a correct interpretation of the phase boundaries as the
boundaries between
plateaux. Such lines are not, strictly speaking, marginal lines, but they
are attracted by the RG fixed point in the middle, which is repulsive in
every direction except along the phase boundaries. These points should
be identified with the middle points of the slopes between plateaux. In this
framework the presence of the $SL(2, Z)$ symmetry is at the basis of the
universality of the critical indices of the transition \cite{nostro} in
agreement with the experimental results. Furthermore
the unstable fixed points seem to be described by a 2D CFT with
central charge $c = 0$
\cite{cardy, lr},
which is
the field theoretical description of the percolative fixed point
\cite{trugman},
and presently we are working on this point (see
also section 5).

\section{The repulsive RG fixed points: a proposal.}

\subsection{General description of the repulsive fixed point.}

In this section we study what happens at the transition between two
distinct phases. This is
done in order to understand the behavior of the system close to the
saddle-point and to show how we can map it onto a percolative model,
according to the analysis given in \cite{trugman}.
Thanks to the modular symmetry $SL ( 2, Z)$ it is sufficient to
study the behavior of the model around only one saddle-point in order to
derive the values of the physical quantities at
all the other points, in agreement with recent experimental results
supporting universality of the critical properties at
 the transition between plateaux \cite{uni}.

Let us give a picture of
the transition between the two condensate phases

\[
( \bar{\nu} , \bar{\mu} ) = ( 1, -1 )
\;\;\; {\rm and} \;\;\;
( \bar{\nu} , \bar{\mu} ) = ( 2, -1 )
\]

from now on to be referred to as (1) and (2)  respectively.

The transition saddle-point that separates them is given by:

\[
\left( \frac{1}{g^* } , \frac{ \theta^*}{2 \pi} \right) =
\left( \frac{1}{2} , \frac{3}{2} \right)
\]

and we will suppose that, once we fix the values of the parameters at the
critical point, the system will be constituted by a ``mixture" of an equal
number of (1) and (2) - particles, let us say
$N$ type-(1) particles and $N$ type-(2) ones.

Starting from such a hypothesis we will construct a mapping of our model to
a percolative transition model. This turns out to be  very
useful a lattice version of the model defined by the action:

\[
S_0 [ \vec{r}^{(1)}_1, \ldots,  \vec{r}^{(1)}_{N}, \;
 \vec{r}^{(2)}_1, \ldots,  \vec{r}^{(2)}_{N} ] =
\]

\[
\frac{g^*}{2} \sum_{ \vec{r}^{(1)}_i \neq \vec{r}^{(1)}_j } ( 1 -
\frac{\theta^*}{2 \pi} )^2 G (\vec{r}^{(1)}_i - \vec{r}^{(1)}_j  )
+ \frac{1}{2 g^* }  G (\vec{r}^{(1)}_i - \vec{r}^{(1)}_j  )
+ \frac{g^*}{2} \sum_{ \vec{r}^{(2)}_i \neq \vec{r}^{(2)}_j } ( 2 -
\frac{\theta^*}{2 \pi} )^2 G (\vec{r}^{(2)}_i - \vec{r}^{(2)}_j )
+  G (\vec{r}^{(2)}_i - \vec{r}^{(2)}_j  )
+
\]

\[
g^* \sum_{\vec{r}^{(1)}_i , \vec{r}^{(2)}_j } ( 1 - \frac{\theta^*}{2 \pi}
) ( 2 - \frac{\theta^*}{2 \pi} ) G ( \vec{r}^{(1)}_i - \vec{r}^{(2)}_j )
+ \frac{1}{g^*}  \sum_{\vec{r}^{(1)}_i , \vec{r}^{(2)}_j }
 G ( \vec{r}^{(1)}_i - \vec{r}^{(2)}_j ) +
\]

\[
i \sum_{ \vec{r}^{(1)}_i , \vec{r}^{(2)}_j } \varphi (
\vec{r}^{(1)}_i - \vec{r}^{(2)}_j ) +
\]

\beq
 \sum_{ \vec{r}^{(1)}_j} \left[ \frac{g^*}{4} ( 1 - \frac{\theta^*}{2 \pi} )^2
+ \frac{1}{4 g^*} \right] ( \vec{r}^{(1)}_j )^2 +
 \sum_{ \vec{r}^{(2)}_j} \left[ \frac{g^*}{4} ( 2 - \frac{\theta^*}{2 \pi} )^2
+ \frac{1}{4 g^*} \right] ( \vec{r}^{(2)}_j )^2
\label{lacr}
\eneq
~~where $ \vec{r}^{(1)}_1, \ldots,  \vec{r}^{(1)}_{N}, \;
 \vec{r}^{(2)}_1, \ldots,  \vec{r}^{(2)}_{N}$ define
the locations of the two
types of particles  and $G( \vr_i - \vr_j )$ and
$\varphi (  \vr_i - \vr_j )$ are the Green functions on the lattice.

We now need to make a second \underline{basic} hypothesis, i.e., that $S_0$
\underline{ does not depend } on the locations of the particles. Such a
statement could be easily proven on a torus,
but for the moment we shall leave it as a guess.

 Next step will be to study what happens
when we ``move'' a little away from the critical point. In order to do
so, let us introduce the fugacities $Y(1)$ and $Y(2)$ for type-(1) and
type-(2) particles respectively.
At the fixed point we have $Y(1) = Y(2) $ and the system may be represented
as a lattice with equal number of particles of both types on its sites.

To move away from the critical point let us suppose that $Y(1) > Y(2)$. In
particular we shall assume that we are changing the value of the parameter
$\frac{\theta}{2 \pi}$ from its critical value, $\frac{\theta^*}{2 \pi}$ to:

\[
\frac{\theta}{2 \pi} = \frac{\theta^*}{2 \pi} + \frac{\delta \theta}{2 \pi}
\]

Once we made such a change there will be two possible configurations:

\begin{itemize}

\item The state to which we shall refer to as  the ``ground state''
, $ | 0 \rangle$,
which is a superposition of states with equal number of particles of
the two types;

\item The state that has $N + 1$ type-(1) particles and $N - 1$ type-(2)
particles obtained after
changing the type of particle lying on a specific
site from (2) to (1).

\end{itemize}

Let us now focus our attention on a given site of the lattice, say 1.

We will suppose that at the critical point site 1 is occupied by a type-(2)
particle. When we move away from the critical point, we can evaluate the
relative probability that 1 is still occupied by a type-(2) particle or
that it is occupied by a type-(1) particle. We find that the weight
corresponding to the configuration with site 1 occupied
 by a type-(1) particle is:

\[
w_1 = k [ Y(1) Y(2) ]^{N} \frac{Y(1)}{Y(2)} \exp \left\{ - S_1 \right\}
\]

where $S_1$ is the action corresponding to a configuration in which
the sites $\vec{r}_1^{(1)}, \ldots, \vec{r}_{N }^{(1)}, \vec{r}_1^{(2)}$
are occupied by
a type-(1) particle while the sites $\vec{r}^{(2)}_1, \ldots,
\vec{r}^{(2)}_{N }$ are occupied by a type-(2) one. The combinatorial
factor $k$ takes into account permutations among identical particles and
is equal to:

\[
k = \frac{ ( 2 N - 1)!}{ N ! ( N - 1)!}
\]

The previous two equations are justified if
for the action $S_1$ we make the same hypothesis as the one  made for the
action $S_0$ (and we shall assume that it may be proven in the same
framework), i.e., that it \underline{does not depend on the locations
of the particles}.

The weight corresponding to a configuration in which  site 1 is
occupied by a type-(2) particle will be:

\[
w_2 = k  ( Y (1) Y(2) )^{N} e^{ - S_0}
\]

Let us finally write down the probabilities $p_1$ for a type-(1) particle
on (1) (``filled site'')  and $p_0$ for a type-(2) particle (``empty
site''):

\[
p_1 = \frac{w_1}{w_1 + w_2 }  \approx \frac{Y(1)}{Y(1) + Y(2)}
\]

\beq
p_0 \approx \frac{Y(2)}{Y ( 1) + Y(2)}
\eneq

Again we made an assumption about the relative values of the action, i.e.
that:

\beq
e^{ S_1^* - S_0^*} \approx 1
\eneq

where the $*$ symbol indicates
that the parameters are to be evaluated at the critical point.

Let us now sketch how our model can be mapped onto a percolative model.

\subsection{Mapping onto a percolative model.}

The ``percolative model'' we shall look at is defined as follows:

\begin{itemize}

\item A 2-dimensional lattice with $2 N$ sites each of which may be
``occupied'' by a particle or ``empty'';

\item A probability $p$ for the site being occupied that is site-independent.

\end{itemize}

It is well-known that such a model presents a phase transition when $p$
increases to its critical value $p_c = .5$. Such a transition appears when
islands of filled sites become a connected cluster that ``percolates''
throughout the lattice. The physical quantity that characterizes such
a transition is the ``correlation length'' $ \xi$ that measures the mean
extension of a big cluster. When $p$ approaches $p_c$, $\xi$ diverges with a
power-law as:

\beq
\xi \sim | p - p_c |^{ - \nu}
\label{pwlaw}
\eneq

where the exponent $\nu$ is equal to $\frac{4}{3} $.

The correspondence with our model
gets traced once we define the probability $p$
of the percolative model as:

\[
p = p_1 = \frac{ Y(1)}{ Y(1) + Y(2)}
\]

This is in agreement with the fact that at the critical point we must have
$p = .5$, corresponding to $Y(1) = Y(2)$.

Since we are able to perform a RG analysis in our model by scaling the
cutoff length $a$ to $ ( 1 + \lambda ) a$ we can relate the index $\nu$
to the scaling exponents of the parameters of our model. In particular
we shall refer to the scaling
properties (at the transition between Hall
plateaux) described in \cite{pru} in terms of
$ \delta \sigma_{H} $. It has been proven there
that, if we rescale length
$L$ by b,
close to the critical point the quantity $\delta \sigma_H = \sigma_H -
\sigma_H^* $  scales according to:

\beq
\delta \sigma_{H } \sim b^{ - \frac{1}{\nu} }
\label{scala}
\eneq

On the other hand
by looking at the mapping introduced above we find:

\[
| p - p_c | \sim | Y ( 1 ) - Y^* ( 1 ) |
\]

Furthermore we shall assume that $Y(1)$ deviates from its critical value
$Y^*(1)$ due to a deviation in the parameter $\frac{\theta}{2 \pi}$ from
its critical value $\frac{\theta^*}{2 \pi} $. Being $Y(1)$
a smooth function of $\frac{\theta}{2 \pi}$ (as it may be easily seen
from the scaling equations given above),
we can write down:

\beq
Y(1) - Y^*(1) \approx \frac{\partial Y^*(1) }{\partial \frac{\theta}{2 \pi}}
\frac{\delta \theta}{2 \pi}
\label{sdev}
\eneq

where $\delta \theta = \theta - \theta^* $ and

\[
\frac{\partial Y^*(1) }{\partial \frac{\theta}{2 \pi}}
\]

is finite.

Finally from eqs.( \ref{pwlaw}, \ref{sdev} ) we get:

\[
\xi ( \theta, \sigma ) \sim \left| \frac{ \delta \theta}{2 \pi}
\right|^{ - \nu}
\]

that is:

\beq
\delta \left( \frac{\theta}{2 \pi} \right) \sim \xi^{ - \frac{1}{ \nu}}
\eneq

The above equation reproduces the scaling properties described in eq.(
\ref{scala} ) once we identify $\delta \left( \frac{\theta}{2 \pi} \right) $
as $\sigma_H - \sigma_H^*$, in agreement with our RG analysis given
in section 3.

\section{Summary, comments and suggestions.}

In this paper we analyzed in detail the long-distance properties of a dual
model recently proposed for describing a Quantum Hall fluid \cite{nostro}.
Even though duality has been lately introduced by several authors for
describing universality, we feel that our proposed model
enjoies two essential features:

\begin{enumerate}

\item
It is general.
It allows for dyon condensation in the vacuum by introducing (a la
$~^{'}$t Hooft) a coupling $\frac{\theta}{2 \pi}$ between the electric and
magnetic charges. Consequently the duality enjoied by the model simply
describes not only the usual symmetry under the exchange of the electric
and magnetic charges but also the symmetry  under
electric (magnetic) charge translation by integer values.

\item The physical interpretation of the two parameters appearing in the
model is \underline{not guessed} but comes out from a detailed (non-linear)
RG analysis describing their flow.

\end{enumerate}

By fixing the values of the parameters at the transition between two
phases, with reasonable hypothesis we traced a mapping between our
description and the classical percolation model. By using the scaling
properties of the $\frac{\theta}{2 \pi}$ parameter we were led to fix the
critical exponent $\nu$ to the value $\nu = \frac{4}{3}$.

We would like to give now some suggestions.

Even though there is no clear experimental result for $\nu$ there is a
general consensus for its value to be $\nu = \frac{7}{3}$ \cite{vkl}.

Can quantum tunneling provide an enhancement of the percolation between
the two phases in such a way that:

\[
\nu = \frac{4}{3} \rightarrow \nu = \frac{7}{3} \;\; ?
\]

How to describe quantum tunneling in our model?

Further is it possible to describe the critical properties of the system in
terms of a 2D CFT? What would be the value of its
central charge?. The percolative picture just presented would suggest
a 2D CFT with central charge $c = 0$ as analytic extension of the unitary
series described by a theory with $ c  = 1 - \frac{6}{m ( m +1 )}$.

In fact it is possible to perturb around the critical point by employing
thermal operators, with an Effective Action given by:

\beq
S^* + \Delta S
\label{efact}
\eneq

~~where $S^*$ describes the system at criticality and the ``perturbation"
$\Delta S$ is given by \cite{cardy}:

\beq
\Delta S = \Delta \left( \frac{ \theta }{2 \pi} \right) \int d^2 z
\Phi_{13} ( z, z^* )
\label{pertur}
\eneq

~~( $\Phi_{13} $ is a thermal operator with conformal dimension $ h =
\frac{5}{8} $ \cite{lr}).

It is straightforward to show that the relation between $\nu$ and $h$ is
given by:

\begin{equation}
\nu = \frac{1}{2 ( 1 - h )}
\end{equation}

obtaining for the critical exponent $\nu$ the value $\nu = \frac{4}{3}$.

An obvious comment is that the above argument is just naive. We think that
it would be interesting to define with care the limit $m \rightarrow
2$ in order to obtain a correct interpretation of the $c = 0$ theory and
then to be able to evaluate the critical index.

At the moment we are workng on this point.

\vspace{1cm}

{\bf Acknowledgements}

We thank R. B.Laughlin for useful discussions.

One of the authors (D. G.) acknowledges support by EU under TMR project
contract FMRX-CT98-0180 and PRA97-QTMD of INFM.

\newpage

{\Large \bf Appendix: $SL(2, Z)$ fixed points and their relation with
the Renormalization Group. }

\vspace{0.5 cm}

A fixed point of $SL(2, Z)$ is a complex number $z_*$ such that it
exists al least one
$A \in SL(2, Z)$ such that $ A z_* = z_* $ and $A \neq 1$.

We can easily prove the following theorem:

``If $z_*$ is a fixed point of $SL(2, Z)$ then $\forall B \in SL(2, Z)$
$B z_*$ is a fixed point too".

To prove such a statement we begin by reminding that there is $A \in
SL(2, Z)$ such that $A z^* = z^*$. If we define $C = BAB^{-1}$,
then we get:

\[
C ( Bz^* )  =Bz^*
\]

which completes the proof.

It is not difficult to prove that
the classification of the fixed points is as follows:

\begin{itemize}

\item $i \infty$, invariant under $\hat{T}$;

\item $i$, invariant under $\hat{S}$;

\item $ \frac{1}{2} + i \frac{\sqrt{3}}{2}$, invariant under
$\hat{T} \hat{S}$.
\end{itemize}

Finally let us notice that, by applying  $SL(2, Z)$ to $i \infty$
we generate the set of rational numbers:

\[
SL(2, Z) ( i \infty ) = Q = \{ \frac{p}{q} \}, q \neq 0
\]

An important property of the fixed $SL(2, Z)$ points is that they are
also fixed RG points (although the vice versa is not necessarily true).
For simplicity we shall assume that
the set of RG fixed points is ``minimal", i.e.,
that the fixed RG points exactly coincide with the fixed points of
$SL(2, Z)$.

Since $\hat{S}^2 = 1$ e $ ( \hat{T} \hat{S} )^3 = 1$,
the points generated by the application of $SL(2, Z)$ to $i$
will be double points of the RG
while the ones generated by $\frac{1}{2} + i \frac{\sqrt{3}}
{2}$ will be triple ones.

The assignment of the RG flux lines  is
chosen in such a way that:

\begin{itemize}

\item it agrees with the ``Physical interpretation" of the parameters;

\item it agrees with the RG flux for the parameters $g$ and $\theta$.

\end{itemize}

Let us now emphasize the physical role of those fixed points:

\begin{itemize}

\item \underline{$C_1$ fixed points} will be the point $i \infty$ and
its images under the group $SL(2, Z)$. They will be identified as
attractive RG fixed points. This means that they are the points
attracting the condensate phases of the system in the large scale
limit. Then all the rational numbers are I.R. fixed points. This is a
relevant consequence of the generalized duality symmetry and is at the
basis of a correct description of the hierarchy.

\item \underline{$C_2$ fixed points} They separate two phases
corresponding to two different Hall condensates. The RG flux lines
assignment shows that these points have the topology of ``saddle
points", i.e., they are attractive in every direction but along the
phase boundary. Then it is evident that they should be identified with
the transition points between plateaux and so they are
localization/delocalization transition points.

\item \underline{$C_3$ fixed points} We will not spend much time in
dealing with them since they will disappear when the modular group
$SL(2, Z)$
gets restricted to $\Gamma(2)$
in order to match the actual physical problem, i.e. the odd denominator
rule \cite{lr}. What we
could tell about them is that they correspond to the opening of new
phases and they are repulsive in each direction.

\end{itemize}

\newpage

{\bf Figure caption: }

It is shown the phase diagram of the model in the
plane $\left( \frac{1}{g}, \frac{\theta}{2 \pi} \right)$.

The attractive IR fixed points are indicated with the symbol $\oplus$,
the saddle points with $\otimes$ and the triple points with $\ominus$.

\end{document}